\documentclass[aps,preprint,showpacs,nofootinbib,12pt]{revtex4}
\usepackage{graphicx}
\usepackage{epsfig}
\usepackage[dvips]{color}

\begin{document}

\title{What do the radiative decays of $X(3872)$ tell us}

\author{ Hong-Wei Ke$^{1}$   \footnote{khw020056@hotmail.com} and
         Xue-Qian Li $^{2}$  \footnote{lixq@nankai.edu.cn}
       }

\affiliation{
  $^{1}$ School of Science, Tianjin University, Tianjin 300072, China \\
  $^{2}$ School of Physics, Nankai University, Tianjin 300071, China
  }

\begin{abstract}
\noindent Since the discovery of $X(3872)$, its structure has been
in ceaseless dispute. The data of $X(3872)\rightarrow
\pi^+\pi^-\pi^0 J/\psi$  suggest that $X(3872)$ may be a high-spin
charmonium-like of $2^{-+}$.  In terms of the light front quark
model (LFQM) we calculate the rates of the radiative decays
$X(3872)\rightarrow J/\psi(\psi')\gamma$ supposing $X(3872)$ to be
a $2^{-+}$ charmonium.  Within this framework, our theoretical
prediction on $\mathcal{BR}(X(3872)\rightarrow\psi(1S)\gamma)$ is
at order of $10^{-3}$ which is slightly lower than  the  Babar's
data but close to the Belle's. Our prediction on
$\mathcal{BR}(X(3872)\rightarrow\psi'\gamma)$ is at order of
$10^{-5}$ if $\psi'$ is a pure 2S state or $10^{-4}$ if $\psi'$ is
a $2S-1D$ mixture, which does not conflict with the upper bound
set by the Belle collaboration, but is much lower than the Babar's
data. Thus if the future measurement decides the branching ratio
of $\mathcal{BR}(X(3872)\rightarrow\psi'\gamma)$ to be much larger
than $10^{-4}$, the  $2^{-+}$ assignment for $X(3872)$ should be
ruled out.

\end{abstract}


\maketitle

\section{introduction}
Very recently, a series of charmonium-like resonances has
successively been observed by various experimental collaborations,
such as $X(3872)$\cite{Choi:2003ue}, $X(3940)$\cite{Abe:2007jn},
$Y(3940)$\cite{Choi:2005}, and several bottomoniun-like states  were
also discovered at the Belle and Babar energy  ranges, such as
$Z(4430)^{\pm}$\cite{Choi:2007wga}, $Z_b$ and $Z_b'$
\cite{Collaboration:2011gj}. It is noted that there is almost no
room in the ground state representations of $O(3)\otimes
SU_f(3)\otimes SU_s(2)$ to accommodate those newly observed
resonances. Not only  their mass spectra, but also their behaviors
of production and decay may hint if they are exotic states such as
hybrids, molecular states, tetraquarks, or just radial and/or
orbital excited charmonia.

It would be an interesting task to determine their inner structures
from both experimental and theoretical aspects. As a matter of fact,
there are different interpretations for these resonances which are
of different constituent compositions from the regular mesons, so
that for clarifying its structure, all the interpretations should be
studied one by one and then we can see if the predictions are in
agreement with data. The most reasonable way is to calculate its
mass and decay widths by assuming its structure in certain
theoretical frameworks, then a comparison of the prediction with
data would confirm or negate  the assumption about the identity of
the resonance.

Among the newly observed resonances, $X(3872)$ which was found a
while ago has caused special interests of experimentalists as well
as theorists
\cite{Mehen:2011ds,Wang:2010ej,Artoisenet:2010va,Narison:2010pd,Nielsen:2010ij,Burns:2010qq}.
Some authors regard it as a molecular
state\cite{Bignamini:2009sk,Voloshin:2003nt,Lee:2009hy}, whereas
other groups consider it as a
tetraquark\cite{Maiani:2004vq,Dubnicka:2010kz,Dubnicka:2011mm}.
Instead, the recent data on $X(3872)\rightarrow \pi^+\pi^-\pi^0
J/\psi$ may hint a possible assignment $^1D_2$ ( i.e.  $2^{-+}$)
for $X(3872)$ which has inspired theoretical
interests\cite{Kalashnikova:2010hv,Brazzi:2011fq,Harada:2010bs,Jia:2010jn}.
The charmonia mass spectrum has been calculated in the potential
model and the theoretical  prediction for a $^1D_2$ charmonium is
about $3.81$GeV\cite{Eichten:1979ms,Buchmuller:1980su} or
$3.84$GeV\cite{Godfrey:1985xj} which is 30-60 MeV lower than the
measured value of $X(3872)$. Because of uncertainty existing in
the calculation of the binding energies of excited states with the
potential model such difference may not be too serious, but the
difference might imply that our understanding on its assignment
might be not completely correct. To determine its structure one
needs more information from other sources, especially its decay
modes. The Belle and Babar collaborations also reported their
measurements on the radiative decays of
$X(3872)$\cite{Abe:2005ix,Aubert:2006aj}. The data given by the
Babar collaboration  \cite{:2008rn} are
$\mathcal{BR}(B^\pm\rightarrow
X(3872)K^\pm)\mathcal{BR}(X(3872)\rightarrow J/\psi\gamma)=(2.8\pm
0.8(stat)\pm0.1(syst))\times 10^{-6},$
$\mathcal{BR}(B^\pm\rightarrow
X(3872)K^\pm)\mathcal{BR}(X(3872)\rightarrow \psi'\gamma)=(9.5\pm
2.7(stat)\pm0.6(syst))\times 10^{-6}$ and
$\frac{\mathcal{BR}(X(3872)\rightarrow
\psi'\gamma)}{\mathcal{BR}(X(3872)\rightarrow
J/\psi\gamma)}=3.4\pm1.4.$ The Belle collaboration also reported
their new result as $\mathcal{BR}(B^\pm\rightarrow
X(3872)K^\pm)\mathcal{BR}(X(3872)\rightarrow
J/\psi\gamma)=(1.78^{+0.48}_{-0.44}\pm 0.12)\times 10^{-6}$ and
set an upper bound $\mathcal{BR}(B^\pm\rightarrow
X(3872)K^\pm)\mathcal{BR}(X(3872)\rightarrow
\psi'\gamma)<3.45\times 10^{-6}$ and
$\frac{\mathcal{BR}(X(3872)\rightarrow
\psi'\gamma)}{\mathcal{BR}(X(3872)\rightarrow J/\psi\gamma)}<2.1$
because they found no evidence for $X(3872)\rightarrow
\psi'\gamma$\cite{Bhardwaj:2011dj}. The data  from the two
collaborations are consistent  on $X(3872)\rightarrow
J/\psi\gamma$ but largely apart on $X(3872)\rightarrow
\psi'\gamma$.  Obviously the further measurement and theoretical
study are badly needed.

In this work we investigate the radiative
decays of $X(3872)$ which is supposed to be a $^1D_2$ charmonium in the light-front
quark model. The results may help us to determine the structure of
$X(3872)$.

The light-front quark model(LFQM) is a relativistic
model\cite{Terentev:1976jk,Chung:1988mu} and in this model
wave functions are manifestly
Lorentz invariant and expressed in terms of the fractions of internal
momenta of the constituents which are independent of the total
hadron momentum. This approach has been applied to study many
processes and thoroughly discussed in literatures
\cite{Jaus:1999zv,Jaus:1989au,Ji:1992yf,Cheng:1996if,Cheng:2003sm,Hwang:2006cua,Ke:2009ed,Ke:2009mn,Li:2010bb,Wei:2009nc,Choi:2007se,Ke:2010vn}.
Generally the results obtained in this framework qualitatively
coincide with the experimental observation on the concerned processes and while taking  the error
ranges into account (both experimental and theoretical), they can
be considered to quantitatively agree with the available data.

To evaluate the transition rate in the LFQM one needs to know the
wave functions of the parent and daughter hadrons. The
wavefunctions for the s-wave and p-wave were given in
Ref.\cite{Cheng:2003sm} and we studied the wavefunctions of the
d-wave in  Ref.\cite{Ke:2011mu} with which we are able to
investigate the transitions involving  s-, p- and d-wave mesons.
Then for obtaining the transition amplitude, one also needs calculate those form factors in terms of the effective Lagrangian. In this work we will deduce the form factors for the radiative
decay of  $2^{-+}\rightarrow 1^{--}$  in the covariant light-front
quark model. With those formulas  we  calculate the rate of the
radiative decays of $X(3872)$ which is supposed to be a pure
$2^{-+}$ meson. Because of the so-called $\rho-\pi$ puzzle, namely
the branching ratio of $\psi'\to\rho\pi$ is extremely small while
$J/\psi\to\rho\pi$ is one of the main decay modes of $J/\psi$, it
is suggested $\psi'$ might not be a pure 2S state ($\psi(2S)$) but
a mixture of $2S$ and $1D$ states ($\psi(2S-1D)$
)\cite{Rosner:2001nm}. This allegation has not been fully
confirmed so far, even though it forms a reasonable interpretation
for the $\rho-\pi$ puzzle and needs more theoretical and
experimental tests indeed. In this work, we consider the two
possibilities and obtain the branching ratio of the decay
$X(3872)\rightarrow \psi'\gamma$ with and without considering the
mixing scenario respectively.

In this work after the introduction we derive the form factors for
the radiative decay of $2^{-+}\rightarrow 1^{--}\gamma$ in the
covariant light-front approach in section II. Then in section III
we present our numerical results about the decay
$X(3872)\rightarrow J/\psi\gamma$ and $X(3872)\rightarrow
\psi'\gamma$. The section IV is devoted to discussions and our conclusion.

\section{the decay of $2^{-+}\rightarrow 1^{--}\gamma$ in the covariant light-front approach}

In Refs.\cite{Jaus:1999zv,Cheng:2003sm} the authors discussed how
to calculate the transition matrix in the covariant light-front
approach. Following their strategy we formulate the matrix element
for $2^{-+}\rightarrow 1^{--}\gamma$. The relative orbital angular
momenta of $2^{-+}$ and $1^{--}$ are $L=2$(d wave) and $L=0$(s
wave) respectively.

Fist let us list the vertex functions for $2^{-+}$\cite{Ke:2011mu}
and $1^{--}$\cite{Cheng:2003sm} states as
\begin{eqnarray}\label{vf9}
iH_{(^{1}D_2)}\gamma_5K_\mu K_\nu,\nonumber\\
iH_V[\gamma_\mu-\frac{1}{W_{V}}(p_1-p_2)_\mu].
\end{eqnarray}
where $V$ represents the $1^{--}$ state and $K=\frac{p_2-p_1}{2}$.
The amplitude of $^1D_2$ state decaying into $1^{--}$ via a photon
emission is written as
\begin{center}
\begin{figure}[htb]
\begin{tabular}{cc}
\scalebox{0.5}{\includegraphics{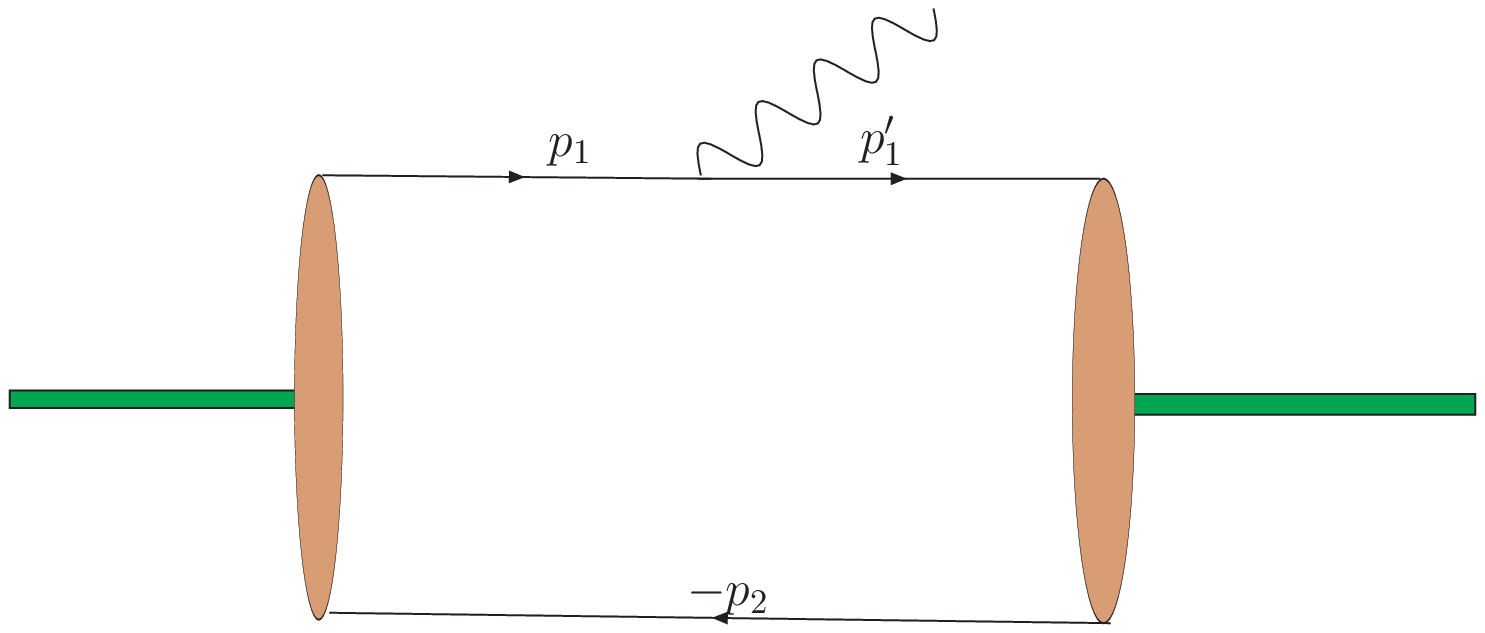}}\\
{\scalebox{0.5}{\includegraphics{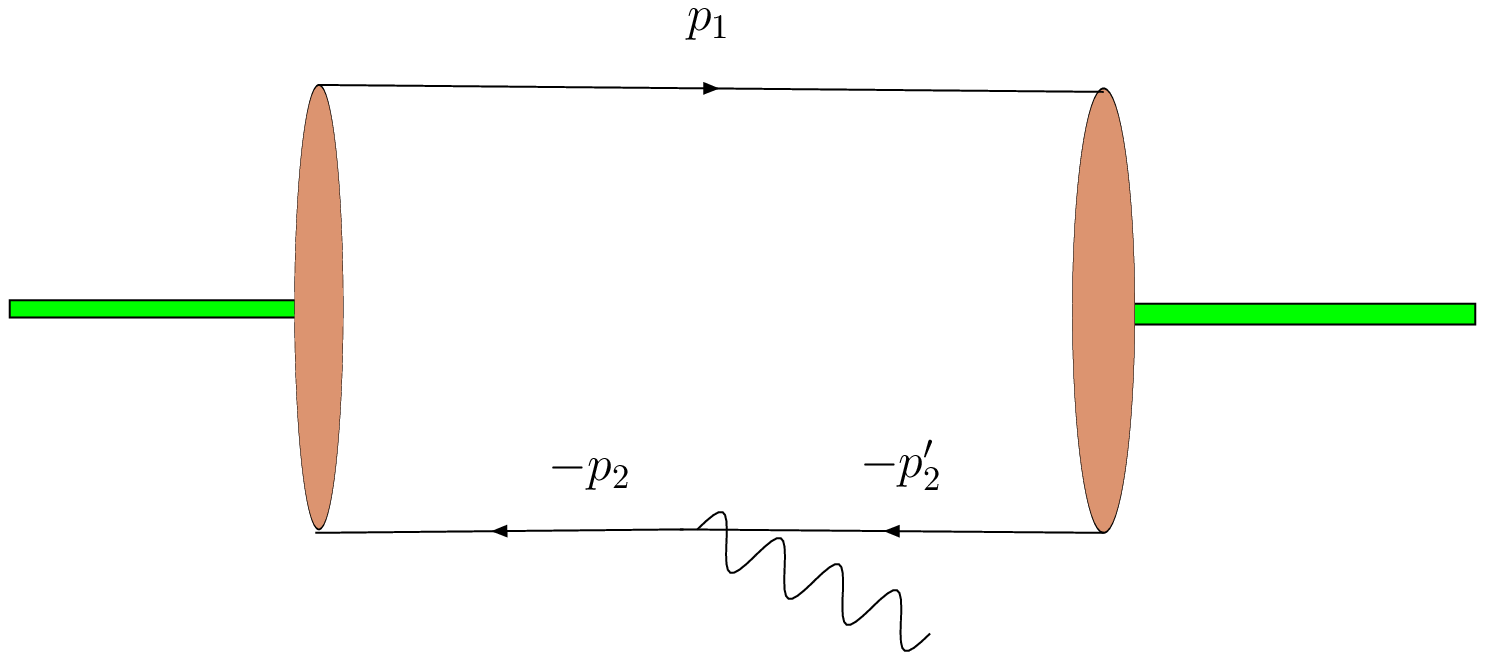}}}
\end{tabular}
\caption{Feynman diagrams depicting the radiative decay in the
light-front quark model.\label{fig:LFQM1}}
\end{figure}
\end{center}

\begin{eqnarray}\label{vf9.1}
\mathcal{A}_\mu&&=-iee_q\frac{N_c}{16\pi^4}\int d^4p_1
[\frac{H_{(^{1}D_2)}H_V}{N_1N_2N_1'}s_{\alpha\beta\mu\nu}^a+
\frac{H_{(^{1}D_2)}H_V}{N_1N_2N_2'}s_{\alpha\beta\mu\nu}^b]{\varepsilon'}^\nu
{\varepsilon}^{\alpha\beta},
\end{eqnarray}
where $$s_{\alpha\beta\mu\nu}^a={\rm
Tr}\{\gamma_5(-\slash\!\!\!p_2+m_2)[\gamma_\nu-\frac{(p_1-p_2)_\nu}
{W_{V}}](\slash\!\!\!p'_1+m_1)\gamma_\mu(\slash\!\!\!p_1+m_1)\}K_\alpha
K_\beta,$$
$$s_{\alpha\beta\mu\nu}^b={\rm
Tr}\{\gamma_5(-\slash\!\!\!p_2+m_2)\gamma_\mu(-\slash\!\!\!p'_2+m_2)[\gamma_\nu-\frac{(p_1-p_2)_\nu}
{W_{V}}](\slash\!\!\!p_1+m_1)\} K_\alpha K_\beta,$$
$N_1=p_1^2-m_1^2+i\epsilon$, $N_1'={p'}_1^2-m_1^2+i\epsilon$,
$N_2=p_2^2-m_2^2+i\epsilon$ and $N_2'={p'}_2^2-m_2^2+i\epsilon$.
The momentum $p_i$  is decomposed as ($p_i^-,p_i^-,{p_i}_\perp$) in
the light-front frame. One needs to integrate over $p_1^-$ by a
contour integration\cite{Jaus:1999zv,Cheng:2003sm}. Here we have
set the relations among all relevant momenta as $
\mathcal{P}=P+P',\,P=p_1+p_2,\, P'=p_1'+p_2'\, q=P-P',\,p_2=p_2'$
for  Fig (a); $p_1=p_1'$ for Fig (b). Then it is easy to find
$s_{\alpha\beta\mu\nu}^a=s_{\alpha\beta\mu\nu}^b$. The integration
contour is closed in the upper plane of the complex $p_1^-$
for the first term in Eq.(2) and in the lower plane for the second
term. Then the first term after the integration turns into
\begin{eqnarray}\label{vf9.2}
\int d^4p_1
\frac{H_{(^{1}D_2)}H_V}{N_1N_2N_1'}s_{\alpha\beta\mu\nu}^a{\varepsilon'}^\nu
{\varepsilon}^{\alpha\beta}\rightarrow-i\pi\int
dx_1d^2p_\perp\frac{h_{(^{1}D_2)}h_V}{x_2 \hat{N_1}\hat{N_1'}}\hat
s_{\alpha\beta\mu\nu}^a\hat{\varepsilon'}^\nu
\hat{\varepsilon}^{\alpha\beta},
\end{eqnarray}
where
\begin{eqnarray*}
h_{(^{1}D_2)}=&&(M^2-M_0^2)\sqrt{\frac{x_1x_2}{N_c}}\frac{1}{\tilde{M_0}\beta^2}\phi(nS),\\
h_V=&&({M'}^2-{M'}_0^2)\sqrt{\frac{x'_1x'_2}{N_c}}\frac{1}{\sqrt{2}\tilde{M'}_0}\phi'(nS)\,\,\,{\rm
for}\,\, ^3S_1,\\
&&({M'}^2-{M'}_0^2)\sqrt{\frac{x'_1x'_2}{N_c}}\frac{1}{\sqrt{2}\tilde{M'}_0}\phi'(nD)\,\,\,{\rm
for}\,\, ^3D_1,\\
\hat{N}_1^{(')}=&&x^{(')}_1({M^{(')}}^2-{M_0^{(')}}^2),\\
\phi'(nD)=&&\frac{\sqrt{6[M_0'^2-(m_1-m_2)^2][M_0'^2-(m_1+m_2)^2]}}
{12\sqrt{5}M_0'^2\beta'^2}\phi'(nS).
\end{eqnarray*} In the formula, $W_V$ and ${p_1}_\mu$, ${p_1}_\mu
{p_1}_\nu$, ${p_1}_\mu {p_1}_\nu {p_1}_\alpha$
 , ${p_1}_\mu {p_1}_\nu {p_1}_\alpha {p_1}_\beta$ in $s_{\alpha\beta\mu\nu}^a$
must be replaced by the appropriate quantities as discussed in
Ref.\cite{Cheng:2003sm} to include the contributions of the zero
modes, for example
\begin{eqnarray*}
&&W_V\rightarrow w_V=M_0+m_1+m_2\,\,\, {\rm for}\,\, \,^3S_1,\\
&&\,\,\,\,\,\,\,\,\,\,\,\,\,\,\,\,\,\,\,\,\,\,\,\,\,\,\,\,\,\,\,\,\,\,
\frac{M_0^2-(m_1+m_2)^2}{2M_0+m_1+m_2}\,\,\,
{\rm for}\,\,\, ^3D_1,\\
&&{p_1}_\mu\rightarrow\frac{x_1}{2}\mathcal{P}_\mu+(\frac{x}{2}-\frac{p_\perp
q_\perp}{q^2})q_\mu,\\
&&\,\,\,\,\,\,\,\,\,\,\,\,\,\,......
\end{eqnarray*}
with $\hat{\varepsilon}^{\alpha\beta}$ and
$\hat{\varepsilon'}^{\nu}$ being identical to
$\varepsilon^{\alpha\beta}$ and $\varepsilon'^{\nu}$ for the
maximally transverse polarized state ($m=\pm J$) and
$s_{\alpha\beta\mu\nu}^a$  changes into  $\hat
s_{\alpha\beta\mu\nu}^a$. More details about the derivation can be
found in Ref.\cite{Cheng:2003sm}. After the replacement, $\hat
s_{\alpha\beta\mu\nu}^a$ is decomposed into
\begin{eqnarray}\label{vf9.2}
\hat
s_{\alpha\beta\mu\nu}^a=&&F_1(\varepsilon_{\beta\mu\rho\omega}\mathcal{P}^\rho
q^\omega
g_{\alpha\nu}+\varepsilon_{\alpha\mu\rho\omega}\mathcal{P}^\rho
q^\omega
g_{\beta\nu})+(F_2+F_3)\varepsilon_{\nu\mu\rho\omega}\mathcal{P}^\rho
q^\omega q_\alpha q_\beta\nonumber\\
&&+F_4(\varepsilon_{\beta\mu\rho\omega}\mathcal{P}^\rho q^\omega
q_\alpha q_\nu+\varepsilon_{\alpha\mu\rho\omega}\mathcal{P}^\rho
q^\omega q_\beta q_\nu),
\end{eqnarray}
with
\begin{eqnarray}
&&F_1=-\frac{4iA_{1}{^{(4)}}}{W_V},\nonumber\\
&&F_2=2iA_{4}{^{(2)}}+2iA_{2}{^{(2)}}-4iA_{3}{^{(2)}},\nonumber\\
&&F_3=\frac{4iA_{4}{^{(4)}}-4iA_{4}{^{(2)}}+8iA_{3}{^{(4)}}}{W_V},\nonumber\\
&&F_4=\frac{4iA_{2}{^{(3)}}-4iA_{3}{^{(2)}}+4iA_{4}{^{(2)}}-4iA_{4}{^{(4)}}}{W_V},
\end{eqnarray}
where $A_{i}{^{(j)}}(i=1\sim4,j=1\sim4)$ are defined in the appendix.

We define the form factors as following
\begin{eqnarray}
&&f_1=\frac{ee_q}{16\pi^3}\int dx_1d^2p_\perp \frac{F_1\phi\phi'}{\sqrt{2}\beta^2 \tilde{M}_0\tilde{M'}_0}(\frac{x_1+x_2}{x_1x_2});\nonumber\\
&&f_2=\frac{ee_q}{16\pi^3}\int dx_1d^2p_\perp \frac{F_2\phi\phi'}{\sqrt{2}\beta^2 \tilde{M}_0\tilde{M'}_0}(\frac{x_1+x_2}{x_1x_2});\nonumber\\
&&f_3=\frac{ee_q}{16\pi^3}\int dx_1d^2p_\perp \frac{F_3\phi\phi'}{\sqrt{2}\beta^2 \tilde{M}_0\tilde{M'}_0}(\frac{x_1+x_2}{x_1x_2});\nonumber\\
&&f_4=\frac{ee_q}{16\pi^3}\int dx_1d^2p_\perp
\frac{F_4\phi\phi'}{\sqrt{2}\beta^2
\tilde{M}_0\tilde{M'}_0}(\frac{x_1+x_2}{x_1x_2}),
\end{eqnarray}
which will be numerically evaluated in next section.

With these form factors   the amplitude is
obtained as
\begin{eqnarray}\label{vf9.1}
\mathcal{A}_\mu&&=
-[f_1(\varepsilon_{\beta\mu\rho\omega}\mathcal{P}^\rho q^\omega
g_{\alpha\nu}+\varepsilon_{\alpha\mu\rho\omega}\mathcal{P}^\rho
q^\omega
g_{\beta\nu})+(f_2+f_3)\varepsilon_{\nu\mu\rho\omega}\mathcal{P}^\rho
q^\omega q_\alpha q_\beta\nonumber\\
&&+f_4(\varepsilon_{\beta\mu\rho\omega}\mathcal{P}^\rho q^\omega
q_\alpha q_\nu+\varepsilon_{\alpha\mu\rho\omega}\mathcal{P}^\rho
q^\omega q_\beta
q_\nu)]{\varepsilon}^\nu{\varepsilon}^{\alpha\beta}.
\end{eqnarray}

\section{Numerical result for $X(3872) \rightarrow J/\psi\gamma$ and $\psi(2S)\gamma$}
With the formulas derived in section II we can calculate the
decay rates for $X(3872) \rightarrow \psi(1S,2S)\gamma$ in the
light-front quark model. In fact  the main task is to evaluate
these form factor $f_1$, $f_2$, $f_3$ and $f_4$ at $q^2=0$.

We set $m_c=1.4$ GeV following Ref.\cite{Cheng:2003sm} at first and
will discuss the dependence of our results on the choice in later
part of the paper. Then we need to fix the parameter $\beta$ in the
wavefunction. Generally in LFQM, we can fix the $\beta$ value by
fitting the decay constant of a vector meson, thus in this work, we
employ $J/\psi$ as the meson. The formula for calculating the decay
constant of a vector meson was given in
Refs.\cite{Jaus:1999zv,Cheng:2003sm}:
\begin{eqnarray}\label{26}
f_V&=&\frac{\sqrt{N_c}}{4\pi^3M}\int dx_1\int
d^2p_\perp\frac{\phi(nS)}{\sqrt{2x(1-x)}\tilde M_0}
\nonumber\\&&\biggl[xM_0^2-m_1(m_1-m_2)-p^2_\perp+\frac{m_1+m_2}{M_0+m_1+m_2}p^2_\perp\biggl],
\end{eqnarray}
where $m_1=m_2=m_c$ and other notations are collected in the
appendix.

Using the data of $J/\psi\rightarrow e^+e^-$\cite{PDG10} we
extract the decay constant of $J/\psi$: $f_{J/\psi}=416\pm 5$ MeV
by which we fix the model parameter $\beta$ as $0.631\pm0.005$GeV.
The masses of $X(3872)$, $J/\psi$ and $\psi(2S)$ are chosen
according to Ref.\cite{PDG10}.

The numerical values of the form factors for $X(3872) \rightarrow
J/\psi\gamma$  are listed in Tab. I. Then we obtain
$\Gamma(X(3872)\rightarrow J/\psi\gamma)=(3.54\pm0.12)\times
10^{-6}$GeV.   The recent measurement sets only an upper bound on
the total width of $X(3872)$, so that we cannot determine accurate
branching ratios for the radiative decays, but only a bound which
is listed in table \ref{tab:decay2}.

Theoretically, Jia $et. al.$\cite{Jia:2010jn}
studied the same transition within the pNRQCD framework where several phenomenological potentials were adopted, in their work the quantum number of $X(3872)$  was assigned as $2^{-+}$.
They obtained $\Gamma(X(3872)\rightarrow
J/\psi\gamma)=(3.11\sim 4.78)\times 10^{-6}$GeV corresponding to
different potential models which is in accordance with our result. For comparing theoretical results
with the data we present the experimental data and theoretical prediction on the branching
ratios $\mathcal{BR}(X(3872) \rightarrow J/\psi\gamma)$
in table \ref{tab:decay2}. The  theoretical
prediction of $\mathcal{BR}(X(3872) \rightarrow J/\psi\gamma)$  is slightly lower than  the Babar's data but quite close to the Belle's data as long as
we suppose $X(3872)$ as a $2^{-+}$ state.
\begin{table}
\caption{the form factors $f_1, f_2, f_3,$ and $f_4$}
\label{tab:decay}
\begin{tabular}{c|c|c|c|c}\hline\hline
 decay mode   &  ~~~~~~$f_1$~~~~~~   &
 ~~~~~~$f_2$~~~~~~ &  ~~~~~~$f_3$~~~~~~ & ~~~~~~$f_4$~~~~~~ \\\hline
 $X(3872) \rightarrow J/\psi\gamma$    & -0.0146 $\pm 0.0002$    &  0.146$\pm0.003$       &  -0.0092$\pm 0.0001$ & 0.0180$\pm$ 0.0001          \\
$X(3872) \rightarrow \psi(2S)\gamma$    & -0.0157$\pm0.0002$    & 0.342$\pm$0.006       &  -0.0144$\pm0.0004 $ &  0.0352$\pm$0.0002  \\
$X(3872) \rightarrow \psi(2S-1D)\gamma$    & -0.0594$\pm0.0004$    & 0.330$\pm$0.005       &  -0.0396$\pm0.0007 $ &  0.0800$\pm$0.005  \\
\hline\hline
\end{tabular}
\end{table}

Now we turn to $X(3872) \rightarrow \psi'\gamma$. We will deal
with the cases where $\psi'$ is regarded  as a pure  $2S$
charmonium $\psi(2S)$\cite{PDG10} or a mixture of 2S and $1D$
charmonium
$\psi(2S-1D)$\cite{Ding:1991vu,Kuang:1989ub,Rosner:2001nm}
respectively and obtain the corresponding numerical results. First
we assume $\psi'$ to be  a pure  2S state to calculate the
branching ratio of  the transition $X(3872) \rightarrow
\psi'\gamma$. Here we need to use the wave function for the radial
excited state. In Ref.\cite{Ke:2010vn} we notice that keeping the
orthogonality among the $nS$ states of heavy quarkonia and
requiring the theoretical predictions on the decay constants of
heavy quarkonia to be in agreement with that obtained from the
data of leptonic decays,  the wavefunction of the radial excited
states nS ($n>$1) should be modified \cite{Ke:2010vn}. In Eq.A(2)
of the appendix we present the explicit form of the modified $2S$
wavefunction. With those wavefunction we obtain the form factors
for $X(3872) \rightarrow \psi(2S)\gamma$ which are listed in Tab.
I.  Now we proceed to evaluate the decay rate of
$X(3872)\rightarrow \psi(2S)\gamma$. We obtain
$\Gamma(X(3872)\rightarrow \psi(2S)\gamma)=(2.44\pm 0.10)\times
10^{-8}$GeV. By the same assumption, the authors of
Ref.\cite{Jia:2010jn} obtained $\Gamma(X(3872)\rightarrow
\psi(2S)\gamma)=1.7\sim 2.9\times 10^{-8}$GeV.

The predicted branching ratio of $\mathcal{BR}(X(3872)
\rightarrow \psi(2S)\gamma)$  presented in table \ref{tab:decay2}
is much lower than  the Babar's
data, but does not contradict to the upper bound
set by the Belle data.

The authors of Ref.\cite{Jia:2010jn} noticed that the mixing of
$\psi(2S)$ and $\psi(1D)$ in $\psi'$ can enhance the rate of
$\Gamma(X(3872) \rightarrow \psi'\gamma)$ remarkably. In this
work, we also evaluate the impact of the mixing in $\psi'$ on
$\Gamma(X(3872) \rightarrow \psi'\gamma)$ with the same mixing
scheme and mixing angle $\theta=12^\circ$ as given in
\cite{Jia:2010jn}. We calculate the form factors which are also
presented in table I and obtain $\Gamma(X(3872) \rightarrow
\psi(2S-1D)\gamma)=(3.65\pm 0.11)\times 10^{-7}$GeV. Our result is
in accordance with that of Ref.\cite{Jia:2010jn} which is within a
range of $(4.9\sim 5.6)\times 10^{-7}$GeV. One can notice that
even though the mixing of 2S and 1D is taken into account the
theoretical prediction  on the branching ratio of
$\Gamma(X(3872)\rightarrow \psi'\gamma)$ is still lower than the
Babar's data by two orders. Moreover, we obtain the ratios
$\frac{\mathcal{BR}(X(3872) \rightarrow
\psi(2S)\gamma)}{\mathcal{BR}(X(3872) \rightarrow
J/\psi\gamma)}=0.069$ and $\frac{\mathcal{BR}(X(3872) \rightarrow
\psi(2S{\rm-}1D)\gamma)}{\mathcal{BR}(X(3872) \rightarrow
J/\psi\gamma)}=0.1$ which obviously contradict to the Babar's
data.

\begin{table}
\caption{Theoretical predictions on the decay widths of the
radiative decay of $X(3872)$, which is regarded as a $^1D_2$
charmonium } \label{tab:decay1}
\begin{tabular}{c|c|c}\hline\hline
   & \cite{Jia:2010jn} & our results
\\\hline
 $\Gamma(X(3872) \rightarrow J/\psi\gamma)$   &$(3.11\sim
4.78)\times 10^{-6}$GeV      &$(3.54\pm0.12)\times
10^{-6}$GeV        \\
$\Gamma(X(3872) \rightarrow \psi(2S)\gamma)$
   & $1.7\sim 2.9\times
10^{-8}$GeV &  $(2.44\pm 0.10)\times
10^{-8}$GeV  \\
$\Gamma(X(3872) \rightarrow \psi(2S-1D)\gamma)$   &$(4.9\sim
5.6)\times 10^{-7}$GeV    & $(3.65\pm 0.11)\times
10^{-7}$GeV \\
\hline\hline
\end{tabular}
\end{table}

\begin{table}
\caption{Experimental data and theoretical predictions on the
branching ratios of the radiative decay of $X(3872)$, which is
regarded as a $^1D_2$ charmonium} \label{tab:decay2}
\begin{tabular}{c|c|c|c|c}\hline\hline
   & \cite{:2008rn}   &
 \cite{Bhardwaj:2011dj} &  \cite{Jia:2010jn} & our results\footnote{ the total width   $\Gamma(X(3872))<2.3$
MeV\cite{PDG10} are used.} \\\hline
 $\mathcal{BR}(X(3872) \rightarrow J/\psi\gamma)$    &$\geq8.75\times 10^{-3}$
   & $\geq5.56\times 10^{-3}$   &  $\geq(1.35\sim2.08)\times 10^{-3}$ &$\geq1.59\times 10^{-3}$        \\
$\mathcal{BR}(X(3872) \rightarrow \psi(2S)\gamma)$
&$\geq2.97\times 10^{-2}$
 & $\sim\,1.01\times 10^{-2}$       & $\geq(7.39\sim12.6)\times 10^{-6}$ &  $\geq10.6\times 10^{-6}$  \\
$\mathcal{BR}(X(3872) \rightarrow \psi(2S-1D)\gamma)$    &    &       & $\geq(2.13\sim2.43)\times 10^{-4}$ &  $\geq1.59\times 10^{-4}$  \\
\hline\hline
\end{tabular}
\end{table}

\begin{center}
\begin{figure}[htb]
\begin{tabular}{cc}
\scalebox{0.9}{\includegraphics{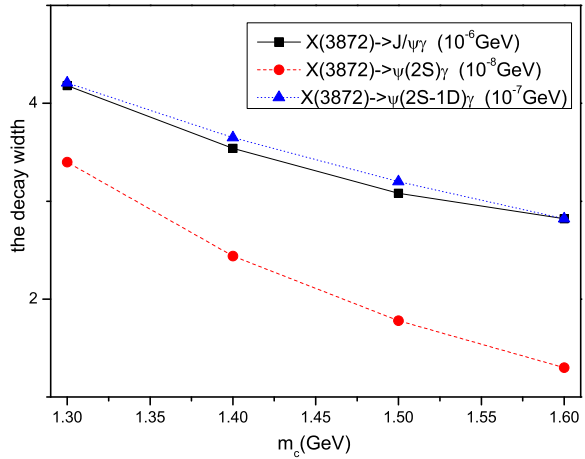}}
\end{tabular}
\caption{The dependence of $\Gamma(X(3872) \rightarrow
J/\psi\gamma)$ and $\Gamma(X(3872) \rightarrow \psi'\gamma)$ on
$m_c$\label{mass}}
\end{figure}
\end{center}

Now we turn to study the dependence of the theoretical predictions
 on $m_c$ for the three radiative decay modes. We let $m_c$ vary within a
reasonable range from 1.3GeV to 1.6GeV and see how the widths of
$\Gamma(X(3872) \rightarrow J/\psi\gamma)$, $\Gamma(X(3872)
\rightarrow \psi(2S)\gamma)$ and $\Gamma(X(3872) \rightarrow
\psi(2S-1D)\gamma)$ change. Indeed, it is noted that when we do so,
we need to refit the $\beta$ parameter for different $m_c$ values
using  Eq.(\ref{26}). The dependence of our results on $m_c$ is
shown in Fig.\ref{mass}. For example, when $m_c$ varies from 1.3GeV
to 1.6GeV the theoretical prediction on $\Gamma(X(3872) \rightarrow
\psi(2S-1D)\gamma)$ changes from $4.21\times10^{-7}$ to
$2.82\times10^{-7}$. It is about 33\% change, but the order of
magnitude remains unchanged. The main goal of the work is to look
for a  criterion to determine the spin-orbit-identity of $X(3872)$.
As aforementioned, the Babar datum is several orders larger than the
prediction, thus the theoretical uncertainty about $m_c$ does not
influence the qualitative conclusion.

\section{Discussion and our conclusion}

Determining the structures of the newly observed resonances is an
interesting and difficult job, but the study is highly rewarding,
because one can eventually identify if they are exotic states.
Retrospecting the history of searching glueballs at the charm
energy range, the results are indeed discouraging. The new
discovery offers an opportunity to recognize the hadrons which are
not existing in the basic representations of $SU(3)$, but as exotic
states or even mixtures of glueball and other states. As a matter
of fact, the charm energy range may be an abundant mine for such
new exotic resonances. However, before one can identify any of
them as an exotic state, he has to thoroughly investigate if it is
an excited state of regular charmonium no matter it is radially or
orbitally excited.  $X(3872)$ is such an unknown resonance. Many
authors consider $X(3872)$ to be a molecular state or tetraquark,
instead, it is also suggested that it may be a $2^{-+}$
charmonium.

In this work we calculate the rates of the radiative decays of
$X(3872)$  in the LFQM. We first deduce the amplitude and
corresponding  form factors in the amplitude of $2^{-+}\rightarrow
1^{--}\gamma$ with the LFQM. Then we suppose the $X(3872)$ is a
pure $2^{-+}$ charmonium and calculate the widths of
$X(3872)\rightarrow J/\psi\gamma$ and $X(3872)\rightarrow
\psi'\gamma$. We obtain $\Gamma(X(3872)\rightarrow
J/\psi\gamma)=(3.54\pm0.12)\times 10^{-6}$GeV,
$\Gamma(X(3872)\rightarrow \psi(2S)\gamma)=(2.44\pm 0.10)\times
10^{-8}$GeV, $\Gamma(X(3872)\rightarrow
\psi(2S-1D)\gamma)=(3.65\pm 0.11)\times 10^{-7}$GeV.

Our theoretical predictions on $\Gamma(X(3872)\rightarrow
\psi'\gamma)$ is obviously inconsistent with the  data of the
Babar collaboration, but that on $\mathcal{BR}(X(3872) \rightarrow
J/\psi\gamma)$ is consistent with data
 within the error
tolerance.

As a matter of the fact, the transition matrix element eventually
depends on an overlapping integral of the wavefunctions of the
initial and final hadrons. Thus the node structure of the
wavefunctions is extremely important. Generally, the state with
the principal quantum number $n$ possesses $n-1$ nodes, thus the
overlapping integral between $X(3872)$ which is supposed to be an
orbital excited state with $n=1$ does not possess a node, and
neither $J/\psi(1S)$, but $\psi(2S)$ does have a node. Therefore,
the integrand in the overlapping integration of $X(3872)$ and
$J/\psi $ is always positive, instead, in the overlapping
integration between wavefunctions of $X(3872)$ and $\psi(2S)$, the
integrand is positive on the left side of the node of the 2S state
and negative on the right side of the node. Thus a cancellation
effect exists. Therefore by a common sense, the transition
$X(3872)\to \psi(2S)+\gamma$ is suppressed and  $B(X(3872)\to
\psi(2S)+\gamma)$ should be smaller than $B(X(3872)\to
J/\psi+\gamma)$.  When the $2S-1D$ mixing for $\psi'$ is
considered, the overlapping integration is decomposed into two
integrations as
$$f_i=\frac{ee_q}{16\pi^3}\int dx_1d^2p_\perp
\frac{F_i}{\sqrt{2}\beta^2
\tilde{M}_0\tilde{M'}_0}(\frac{x_1+x_2}{x_1x_2})\phi[\cos\theta\phi'(2S)-\sin\theta\phi'(1D)]$$
where $\phi'(2S)$, $\phi'(1D)$, $\phi$ are the radial parts of the
wavefunctions  for $\psi(2S)$, $\psi(1D)$ which are the
ingredients in $\psi'$, and $X(3872)$, $\theta$ is the mixing
angle of $2S$ and $1D$ states in $\psi'$. There is a node in
$\phi'(2S)$, but not in $\phi'(1D)$, so the first overlapping
integration is suppressed as discussed before, but not for the
second one. Therefore, the predicted branching ratio is enhanced
in comparison with the pure $2S$ state assignment for $\psi'$.

The data of the Babar and Belle collaborations on $X(3872)\to
\psi'+\gamma$ are obviously different, so that more accurate
measurements are needed.

Our conclusion is that if the branching ratio of $X(3872)\to
\psi'+\gamma$ is indeed as large as the Babar group measured, the
$2^{-+}$  charmonium assignment of $X(3872)$ should be ruled out.

\section*{Acknowledgments}

This work is supported by the National Natural Science Foundation
of China (NNSFC) under the contract No. 11075079 and No. 11005079;
the Special Grant for the Ph.D. program of Ministry of Eduction of
P.R. China No. 20100032120065.
\appendix

\section{Notations}

Here we list some variables appearing in the context.  The
incoming  meson in Fig. \ref{fig:LFQM1} has the momentum
$P=p_1+p_2$ where $p_1$ and $p_2$ are the momenta of the off-shell
quark and antiquark and
\begin{eqnarray}\label{app1}
&& p_1^+=x_1P^+, \qquad ~~~~~~p_2^+=x_2P^+, \nonumber\\
&& p_{1\perp}=x_1P_{\perp}+p_\perp, \qquad
 p_{2\perp}=x_2P_{\perp}-p_\perp,
 \end{eqnarray}
with  $x_i$ and $p_\perp$ are internal variables and $x_1+x_2=1$.

The variables $M_0$, $\tilde {M_0}$ and $\hat{N_1}$ are defined as
\begin{eqnarray}\label{app2}
&&M_0^2=\frac{p^2_\perp+m^2_1}{x_1}+\frac{p^2_\perp+m^2_2}{x_2},\nonumber\\&&
\tilde {M_0}=\sqrt{M_0^2-(m_1-m_2)^2},\nonumber\\&&
\phi(1S)=4(\frac{\pi}{\beta^2})^{3/4}\sqrt{\frac{dp_z}{dx_2}}{\rm
 exp}(-\frac{p^2_z+p^2_\perp}{2\beta^2}),\nonumber\\&&
 \phi(2S)=4\Big(\frac{\pi}{\beta^2}\Big)^{3/4}\sqrt{\frac{\partial
p_z}{\partial
x_2}}{\exp}\Big(-\frac{{2}^\delta}{2}\frac{p^2_z+p^2_\perp}{\beta^2}\Big)
\Big(a_2 -b_2\frac{p^2_z+p^2_\perp}{\beta^2}\Big).
 \end{eqnarray}
with $p_z=\frac{x_2M_0}{2}-\frac{m_2^2+p^2_\perp}{2x_2M_0}$,
$\delta=1/1.82$, $a_2=1.88684$ and $b_2=1.54943$.

The $A_{ij}(i=1\sim 4, j=1\sim4)$ are
\begin{eqnarray}\label{app2}
&&A_{1}{^{(1)}}=\frac{x_1}{2},\,\,\,
A_{2}{^{(1)}}=A_{1}{^{(1)}}-\frac{p_{\perp}\cdot
q_{\perp}}{q^2},\,\,\,A_{1}{^{(2)}}=-p_{\perp}^2-\frac{(p_{\perp}\cdot
q_{\perp})^2}{q^2},\nonumber\\&&
A_{2}{^{(2)}}=({A_{1}{^{(1)}}})^2,\,\,\,A_{3}{^{(2)}}=A_{1}{^{(2)}}A_{2}{^{(2)}},\,\,\,A_{4}{^{(2)}}=(A_{2}{^{(1)}})^2-\frac{A_{1}{^{(2)}}}{q^2},
\nonumber\\&&A_{1}{^{(3)}}=A_{1}{^{(1)}}A_{12},\,\,\,A_{2}{^{(3)}}=A_{2}{^{(1)}}A_{1}{^{(2)}},\,\,\,
A_{3}{^{(3)}}=A_{1}{^{(1)}}A_{2}{^{(2)}},\,\,\,A_{4}{^{(3)}}=A_{2}{^{(1)}}A_{2}{^{(2)}},
\nonumber\\&&
A_{1}{^{(4)}}=\frac{(A_{1}{^{(2)}})^2}{3},\,\,\,A_{2}{^{(4)}}=A_{1}{^{(1)}}A_{1}{^{(3)}},\,\,\,A_{3}{^{(4)}}=A_{1}{^{(1)}}A_{2}{^{(3)}},\,\,\,A_{4}{^{(4)}}=A_{2}{^{(1)}}
A_{1}{^{(3)}}-\frac{A_{1}{^{(4)}}}{q^2}.
 \end{eqnarray}

\end{document}